\documentclass[12pt,reqno,aasms4]{article}
\usepackage{a4}
\usepackage{amsmath, amssymb}              % for mathematical symbols etc
\usepackage{graphicx}                      % for graphics
\numberwithin{equation}{section}

\renewcommand{\phi}{\varphi}                % shortcut

\newcommand{\e}{\mathrm e}                  % Euler number

\begin{document}

\title{Global Dynamics in the Singular Logarithmic Potential}

\author{Cristina Stoica, Andreea Font}

\footnotetext[1]{e-mail: C.Stoica@surrey.ac.uk. Address: Department of Mathematics and Statistics, School of ECM, University of Surrey, Guilford, Surrey, GU2 7XH, UK}
\footnotetext[2]{e-mail: afont@beluga.phys.uvic.ca. Address: Department of Physics and Astronomy, University of Victoria, Victoria, BC, V8P 1A1, Canada}

\begin{abstract} 
 
We present an analytical description of the motion in the singular logarithmic potential of the form $\displaystyle{\Phi =\ln \sqrt{x_1^2/b^2+x_2^2},}$ a potential which plays an important role in the modeling of triaxial systems (like elliptical galaxies) or bars in the centers of galaxy disks. In order to obtain information about the motion near the singularity, we resort to McGehee -type transformations and regularize the vector field. In the axis-symmetric case $(b=1),$ we offer a complete description the global dynamics. In the non axis-symmetric case $(b<1),$ we prove that all orbits, with the exception of  a negligible set, are centrophobic and retrieve numerically  partial aspects of the orbital structure.

\vspace{0.3cm} {\textbf{Key words}}: non axis-symmetric logarithmic
potential, regularization, dynamics near the singularity.
\end{abstract}

\maketitle

\section{Introduction}

The non axis-symmetric logarithmic potential plays an important role
in galaxy dynamics. In the three-dimensional space, the potential:

\begin{equation}
\Phi = \frac{1}{2} \ln({R_{c}}^{2} + \frac{{x_{1}}^{2}}{b^{2}}
+ {x_{2}}^{2})
\end{equation}

\noindent (where $(x_{1},x_{2})$ are the usual cylindrical coordinates $x_{1}=z,$ $ x_{2}=R$) models an elliptical galaxy  with a dense core of radius $R_{c}$ and with the additional property of having a flat rotation curve at large
radii \cite{BT}. In the 2D space, the logarithmic potential can
describe other non axis-symmetric components of galaxies, such
as bars in the centers of galaxy disks.

The study of the orbital structure of the logarithmic potential was initially motivated by 
the need to construct self-consistent models of galaxies\cite{Sc}.
Numerical experiments have proven to be very useful in revealing the rich orbital structure of this potential, including the major orbit 
families, the resonances and the stochastic orbits. For example, in the axis-symmetric 
case ($b = 1$), the logarithmic potential has been shown to admit only loop orbits, which are regular and avoid the origin. The non axis-symmetric ($b < 1 $) potential admits two major families of orbits: box (for $\rho = \sqrt{R^2 +z^2} \ll R_{c})$ and loop orbits (for $\rho \gg R_{c}$) (\cite{BS}, \cite{R}). Note that this behavior can be also be retrieved from the two simple analytical approximations of the potential, one as a sum of two oscillators very close to the origin and the
other one as $\sim ln{(\rho)}$ at large distances \cite{BT}.

An interesting change in the orbital behavior has been discovered when the
potential becomes singular. In their numerical study, Miralda-Escud\'e 
$\&$ Schwarzschild \cite{MS} have found that as $R_{c}\rightarrow 0$, 
a larger fraction of regular box orbits becomes irregular or ``box-like''
(i.e. they will admit fewer integrals of motion than the number of
spatial dimensions), with the end result that, in the limit $R_{c}
= 0,$ all box orbits are irregular. The general interpretation of this result is that 
the scattering by the singularity renders the box orbits unstable, a similar 
behavior to that observed in systems which contain a central black hole \cite{GB}.
In addition, the singular logarithmic potential admits several families of minor orbits, 
{\it i.e.,} resonances in terms of the $x_{2}$:$x_{1}$ frequency ratio: the banana (2:1), 
 fish (3:2) and pretzel (4:3) orbits \cite{MS}.
Numerical studies have also revealed the existence of some stochastic orbits in a narrow region near the singularity (\cite{MS}, \cite{TT}). This has led some to suggest a link between the scattering by the singularity and the transition to chaos \cite{KC}, although no rigorous proof has been given to date in support of this hypothesis.

The present investigation is an analytical approach to the study of the orbital dynamics (including the behavior near the singularity) in the case of the singular logarithmic potential:

\begin{equation}
\Phi = \frac{1}{2} \ln({\frac{{x_{1}}^{2}}{b^{2}}+ {x_{2}}^{2}}),
\label{log1}
\end{equation}

\noindent in both the axis-symmetric and non axis-symmetric cases. The aim of our work is not only to complement the previous numerical studies performed on this subject, but also to offer a theoretical basis for interpreting their results. 

The main difficulty in investigating the system analytically is due to the presence of the singularity in the origin, which creates a discontinuity in the equations of motion. 
This problem motivates the introduction of a change of coordinates that
regularizes the equations of motion. For this, we resort to McGehee-type transformations \cite{M}, a technique that is frequently used in celestial mechanics for the study of singularities in the n-body problem $(n=1,2,3, ...).$ The underlying idea behind this technique is to transform the equations of motions and the time, such that the singularity is "blown-up" into a non-trivial manifold (in our case a torus). By studying the characteristics of the flow on this manifold, one can extrapolate the information (by continuity with respect to the initial data) about the orbital dynamics around the singularity \cite{St}. 

The paper is organized as follows: Sections 2 and 3 provide a brief description of our system, in terms of the equations of motion and the conservation of energy. In Section 4, we remove the singularity by regularizing the equations of motion. In Section 5, we provide a description of the collision manifold and the zero velocity manifold, two abstract surfaces on which we can visualize the properties of the flow close to the singularity and at the maximum distance from the source allowed for a given energy, respectively. Section 6 presents the complete global dynamics in the axis-symmetric case ($b=1$). In Section 7, we extend the analysis for the case of the non axis-symmetric potential ($b \neq 1$), in the restricted limit in which the anysotropy is small. In this case, we prove theoretically that the majority of orbits in this case are centrophobic (that is, they avoid the origin) $-$ a result that has been originally discovered in the numerical study of Miralda-Escud\'e $\&$ Schwarzschild \cite{MS} $-$ and we discuss the orbital structure in terms of orbits which preserve or change the sign of their angular momentum. Finally, in Section 8, we show how several families of resonances can be retrieved through the numerical integration of the new system of equations.

\section{The Equations of Motion}

The non axis-symmetric logarithmic problem is a one-parameter Hamiltonian
system with two degrees of freedom. The anisotropic logarithmic
singular potential (\ref{log1}) determines a conservative system
with a preserved Hamiltonian given by:
\begin{equation}
H(x_{1},x_{2},y_{1},y_{2})=\frac{1}{2}({y_{1}}^{2} + {y_{2}}^{2})
+ \ln(\sqrt{\frac{{x_{1}}^{2}}{b^{2}} + {x_{2}}^{2}}) \label{log2}
\end{equation}
where $(x_{1},x_{2}) \in \mathbb{R}_+^2 - \{(0,0)\}$ are the
generalized coordinates and $(y_{1},y_{2}) \in \mathbb{R}$ are the
momenta.

We start by writing the equations of motion in a form that
contains the anisotropy in the kinetic term rather than in the
potential.  Thus, we substitute $q_{1} = \frac{x_{1}}{b^{2}},$
$q_{2} = x_{2},$ and $\displaystyle{p_{1}=y_{1}/b^2},$
$\displaystyle{p_{2}=y_{2}}.$ Introducing the (standard) notations
for the generalized vector coordinate ${\bf q}=(q_{1}, q_{2}),$
the generalized momenta ${\bf p}=(p_{1}, p_{2})$, the anisotropy
parameter ${\displaystyle \mu =
\frac{1}{b^{2}}}$ and the mass matrix $$M = \left(\begin{array}{cc} \mu & 0 \\0 & 1 \\
\end{array}\right),$$ the Hamiltonian writes:

\begin{equation}
H({\bf{q}, \bf{p}})=\frac{1}{2}{\bf p}^T M {\bf p} + \ln |\bf{q}|
 \label{log3}
\end{equation}
and the anisotropic logarithmic problem is given as the first
order system of ordinary differential equations:

\begin{equation}
\begin{cases} \dot{\bf{q}} = M {\bf{p}} \cr
\dot{\bf{p}} = - \frac{\bf{q}}{\left|{\bf{q}} \right| ^{3}}
\end{cases}
\label{log4}
\end{equation}

When $\mu=1,$ the above system describes the motion in the
axis-symmetric logarithmic potential. From the point of view of
Hamiltonian mechanics this case is completely integrable, as we
have the two integrals of motion given by the conservation of the
total energy and the total angular momentum. However, we point out 
that for $|{\bf q}|\rightarrow 0$ the dynamical
behavior becomes unknown since the vector field $({\bf \dot q}, {\bf \dot p})$ ceases to exist.

 Let K be the kinetic energy,

\begin{equation} K = \frac{1}{2} {\bf{p}}^{T} M {\bf{p}} ,\end{equation} and  V the
potential,
\begin{equation}V =  \ln{\left| {\bf{q}} \right|}.\end{equation}
The total energy $E$ is then:

\begin{equation} H({\bf{q}},{\bf p}) = K + V.\end{equation}
Since the system (\ref{log4}) is Hamiltonian the total energy is
conserved. That is $H({\bf{q}},{\bf p})$ is constant
 along the solution curves of (\ref{log4}) and consequently the
 level sets of $H({\bf{q}},{\bf p})$ are invariant under the flow. If $h$ is a certain
 energy constant (i.e. $H({\bf{q}},{\bf p})=h=$constant) then the level set
 $H({\bf{q}},{\bf p})^{-1}(h)$ is a three-dimensional surface usually called an
 {\it energy surface,} which we denote $\Sigma_{h}.$

\section{Topological Description of the Energy Surfaces}

We want to consider in more detail the topology of various
surfaces $\Sigma_{h},$ including the orbits near the singularity.
Let us fix $h \in \mathbb{R}.$ Using a technique similar to
McGehee's \cite{M} we introduce the change of variables:

\begin{equation} \begin{cases}
{\bf{q}} = r e^{-\frac{1}{r^{2}}} {\bf{s}} \\
{\bf{p}} = \frac{1}{r} {\bf{u}}
\end{cases}
\end{equation} where $r > 0$, $\bf{s}$ is a point on the unit
circle $\mathbb{S}^1$ and $\bf{u} \in \mathbb{R}^{2}$. Our transformation
is a diffeomorphism from $\mathbb {R}^2 \setminus \{(0,0)\} \times
\mathbb{R}^2$ to $(0,\infty) \times \mathbb{S}^1 \times
\mathbb{R}^2$ (where $\mathbb{S}^1$ is the unit circle) and can be
understood as a passing to some unorthodox  kind of polar
coordinates.

The system (\ref{log4}) becomes:

\begin{equation}
\begin{cases}
\displaystyle{
\dot{r} = \frac{r}{r^2+2} \e^{\frac{1}{r^{2}}} {\bf{s}}^{T} M {\bf{u}}} \\
\displaystyle{\dot{\bf{s}} = \frac{1}{r^{2}} \e^{\frac{1}{r^{2}}}
 \Big [M {\bf{u}}
 - ({\bf{s}}^{T} M {\bf{u}}) \cdot \bf{s} \Big ]} \\
\displaystyle{\dot{\bf{u}} = \e^{\frac{1}{r^{2}}} \Big [
\frac{1}{r+2}
 ({\bf{s}}^{T} M {\bf{u}}) \cdot {\bf{u}} - \bf{s} \Big ]}\\
\label{log5}
\end{cases}
\end{equation}
and the conservation of energy transforms to:

\begin{equation}
\frac{1}{2} {\bf{u}}^{T} M {\bf{u}} + r^{2} \ln{r} - 1 = h \cdot
r^{2}. \label{log6}
\end{equation}

Notice that the new system (\ref{log5}) is analytic on the open
manifold $(0,\infty) \times \mathbb{S}^1 \times \mathbb{R}^2$ and
that the regions of motion are constrained by the energy relation (\ref{log6}).
More precisely, since the kinetic term in (\ref{log6}) is
positive, we have that for a fixed level of energy $h$:

\begin{equation}
h r^{2}- r^{2} \ln{r} + 1 \geq 0.
 \label{log7}
\end{equation}
Solving the above relation, it follows that $0 <r< R_{max}$ with $R_{max}
=R_{max}(h).$ In other words the motion is always bounded for any
fixed level of energy $h.$  Also, we can say that the energy
surface $\Sigma_{h}$ projects onto a disc of radius $R_{max}$ in
$\mathbb{R}^2 \setminus (0,0).$ Since

\begin{equation}
\frac{1}{2} {\bf{u}}^{T} M {\bf{u}} = h r^{2}- r^{2} \ln{r} + 1
\label {log8}
\end{equation}
it follows that along the $r=R_{max}$ boundary we have  ${\bf
u}=\bf {0},$ that is the kinetic term cancels. For this reason the
curve
\begin{equation}
\begin{cases}
r=R_{max} \cr
 \bf u= \bf 0
\end{cases}
\label {log9}
\end{equation}
is called the {\it oval of zero velocity} in $\Sigma_{h}.$ In physical 
space, this curve represents the outmost boundary of the orbital structure 
allowed for a given energy. We will denote this curve by $Z.$

\vspace{0.3cm} \noindent PROPOSITION 3.1 For any fixed level of
energy $h$ the energy surface $\Sigma_{h}$ is diffeomorphic to an
open solid torus (i.e. a solid torus minus its boundary).

\vspace{0.3cm} \noindent $Proof$ The proof follows closely the
proof of Proposition 1.1 in \cite{D}. Let $B$ be the interior of
the ellipse
$$\frac{1}{2} {\bf{u}}^{T} M {\bf{u}}=1$$ in the plane. Clearly $B$ is topologically equivalent to $\mathbb{S}^1 \times (0, R_{max}]$ ($r=0$
corresponds to the missing boundary of the ellipse). Then we define $ F:
\Sigma_{h}\longrightarrow \mathbb{S}^1 \times B$ by
$$ F(r,{\bf s},{\bf u})= ({\bf s},{\bf u} ).$$ $F$ is just the
required diffeomorphism.
\begin{flushright}
%$\Box$
\end{flushright}

\section{Regularization of the Vector Field}

\noindent The objective of this section is twofold: to extend the
energy relation to $r=0,$ and to regularize system (\ref{log6})
such that the new vector field becomes differentiable over the entire
interval  $r \in [0, R_{max}].$

The energy relation (\ref{log6}) is ill defined at the singularity
$r=0.$ However, the function $r^2 \ln r$ can be continuously
extended for $r=0$ by:

\begin{equation}
f(r) =
\begin{cases}
r^{2} \ln{r} & \text{if $r > 0$ ;}\\
0 & \text{if $r = 0$}
\end{cases}
\label{log12}
\end{equation}
The extended function (\ref{log12}) is also differentiable over its domain.
Therefore we are able to extend the energy manifold by
\begin{equation}
\frac{1}{2} {\bf{u}}^{T} M {\bf{u}} + f(r) - 1 = h r^{2},
\label{energy}
\end{equation} for all $r \in [ 0, R_{max}].$

 Now we will introduce a sequence of transformations of the system (\ref{log5})
such that the new system will have no singularity at $r=0.$ Following the 
technique introduced by McGehee \cite{M}, we want to paste an invariant manifold
onto the phase space, such that we close the open solid torus that bounds
the motion, by including its boundary $r=0.$ Also, in order to preserve 
the continuity of the flow with respect to the initial data, 
we have to ensure that the transformed system has a differentiable vector
field.

We implement a  change of the time variable through
$ \displaystyle{d\sigma = - \e^{\frac{1}{r^{2}}}  r^{2} dt}$. This will have the effect of decreasing the rate of the time intervals near the singularity. Expressed in the new time derivative $\frac{d}{d\sigma},$ the system (\ref{log5}) becomes:
\begin{equation} \begin{cases}
\dot{r} = - \frac{r^{3}}{r^2+2} {\bf{s}}^{T} M {\bf{u}} \\
\dot{\bf{s}} =  ({\bf{s}}^{T} M {\bf{u}}) \cdot {\bf{s}} - M {\bf{u}} \\
\dot{\bf{u}} =  r^{2} [{\bf{s}} - \frac{1}{r^2+2} ({\bf{s}}^{T} M {\bf{u}}) \cdot {\bf{u}} ]\\
\end{cases}
\label{presystem}
\end{equation}
We note that by the above sequence of reparametrizations we have
obtained an analytic vector field for $(r, {\bf s},{\bf u}) \in
[0, R_{max}] \times \mathbb{S}^1 \times \mathbb{R}^2,$ which is
coupled with a differentiable integral relation given by
(\ref{energy}).

It follows from (\ref{energy}) that each energy surface
 $\Sigma_{h}$ meets the boundary $r=0$ along a submanifold given
 by
\begin{equation} \begin{cases}
\displaystyle{\frac{1}{2}{\bf u}^T M {\bf u}=1} \\
{\bf s} \hspace{0.2cm}arbitrary.\\
\end{cases}
\label{4.4}
\end{equation}
This manifold, let us call it $\Lambda,$ is diffeomorphic with a
two-dimensional torus, which we shall call the {\it collision
manifold}. This fictitious torus has no meaning in the physical space, 
since the motion ceases to exist in the origin. However, the behavior 
of the orbits near the singularity can be extrapolated from similar properties
of the flow on the collision manifold. In this sense, we regard the differential 
system of equations as a vector field on the manifold, its solutions representing
 the flow. 

Note that $\Lambda$ is independent of the total energy
and therefore we can say that the time transformations we have
applied have the effect of pasting an invariant boundary onto each
$\Sigma_{h}.$ Over this boundary the vector field is given by
\begin{equation} \begin{cases}
\displaystyle{\dot{\bf{s}} = ({\bf{s}}^{T} M {\bf{u}}) \cdot {\bf{s}} - M {\bf{u}} }\\
\displaystyle{\dot{\bf{u}} = {\bf 0}}\\
\end{cases}
\label{4.5}
\end{equation}

The solid and now compact torus $\Sigma_{h}$ bounds
all the orbits and by investigating its properties, one can obtain a global
picture of the motion, including the motion near the singularity.
We emphasize that the orbits of system~(\ref{log4}) are the same
as the orbits of system~(\ref{presystem}), only the
parametrization is different. Therefore, any results concerning the solutions
of the first system can be seen as results for the solutions of
the second, as long as one is aware of the fact that the rate at
which solutions move along the orbits is different.

Using the energy integral we can further reduce the dimension of the system. 
For this, we express the coordinates $({\bf{s}},{\bf{u}})$ in terms of the new
angle coordinates $\theta$ and $\psi$:

\begin{equation}
 \begin{cases}
{\bf{s}} = (\cos{\theta}, \sin{\theta})\\
\displaystyle{{\bf{u}} = \sqrt{2 (h r^{2} - f(r) + 1)}
(\frac{1}{\sqrt{\mu}}\cos{\psi}, \sin{\psi})}
\end{cases}
\label{reduce}
\end{equation}

\noindent Also, in order to further simplify the system, we perform a similar time parametrization which incorporates a part of the radial dependence in a new time variable:

\begin{equation}\displaystyle{ds = \sqrt{2 (h r^{2} - f(r) + 1)}
d\sigma.} \label{log11} \end{equation} In these variables, the system  (\ref{presystem}) becomes a first order system
for $(r, \theta, \psi) \in [0, R_{max}] \times \mathbb{S}^1 \times
\mathbb{S}^1$ with a differentiable vector field:
\begin{equation} \begin{cases}
\displaystyle{\dot{r} = - \Big (\frac{r^{3}}{r^2+2}\Big) 2(h r^{2} - f(r) +1) (\sqrt{\mu} \cos{\theta} \cos{\psi} + \sin{\theta} \sin{\psi})}\\
\displaystyle{\dot{\theta} = 2 (h r^{2} - f(r) + 1) (\sqrt{\mu} \cos{\psi} \sin{\theta} - \sin{\psi} \cos{\theta})}\\
\displaystyle{\dot{\psi} = r^{2}(\cos{\psi} \sin{\theta} -
\sqrt{\mu} \sin{\psi} \cos{\theta}).}
\end{cases}
\label{log15}
\end{equation}

Before discussing different aspects of the dynamics of the above
system, we make two more observations:

\noindent OBSERVATION (1) Up to this point, no reference was 
made to the angular momentum integral. It is well known that in
the axis-symmetric case $(\mu=1),$ the singular logarithmic system admits,
besides the total energy, a second conserved quantity, namely the
angular momentum $C:=q_2p_1-q_1p_2.$ In terms of $(r, \theta,
\psi)$ this reads:
\begin{equation}
\displaystyle{ C(s) =  e^{-1/r^2}2[hr^2 - f(r) +1]
\big[\sqrt{\mu}\sin \theta \cos \psi - \cos\theta \sin \psi
\big],} \label{moment}
\end{equation}where $s$ is the final time variable introduced in (\ref{log11}).

Like the energy relation, $C(s)$ is ill defined at the singularity. In an analogous way, we extend the function $e^{-1/r^{2}}$ continuously at $r=0$ by defining the function:
\begin{equation}
g(r) =
\begin{cases}
e^{-1/r^2} & \text{if $r > 0$ }\\
0 & \text{if $r = 0$}
\end{cases}
\label{g}
\end{equation}
The angular momentum is now well defined and differentiable for
all $r \geq 0.$  For the axis-symmetric case ($\mu=1$), the law of conservation of angular momentum ensures that:
\begin{equation}
\displaystyle{ \frac{dC}{ds}=0}
\end{equation}
and therefore we have the integral relation:
\begin{equation}
\displaystyle{C(s)= g(r)2[hr^2 - f(r) +1] \sin (\theta -\psi) =
const. } \label{momentisotropic}
\end{equation}
 In the anisotropic case ($\mu \neq 1$) the above symmetry is lost. The
 anisotropy is responsible for the much more
complicated dynamics and, eventually, for the existence of the chaotic
motion. For later purposes, we write the variation of angular
momentum
\begin{equation}
\displaystyle{\frac{dC}{dt}=(1-\mu)p_1p_2,} \label{4.11}
\end{equation}
in terms of the new variables and time:
\begin{equation}
\displaystyle{\frac{d}{ds}\Big(e^{-1/r^2} \dot \theta
\Big)=\frac{1-\mu }{\sqrt \mu}e^{-1/r^2} 2[hr^2 - f(r) +1]^{3/2}
\sin \psi \cos \psi.} \label{angular}
\end{equation}

\vspace{0.3cm} \noindent OBSERVATION (2) The orbital dynamics of
the system (\ref{log15}) does not depend on the level of energy $h$.
More precisely, each different but fixed $h$ gives rise to the
same qualitative phase portrait. The only change is in the value of 
 $R_{max},$ as $R_{max}$ depends directly on $h.$ Therefore,
without losing generality, and in order to simplify the
calculations, we choose from now on to work with $h=0$.

\section{The Collision Manifold and the Zero Velocity Manifold}

The flow on the invariant collision manifold $\Lambda$ is given by imposing
the restriction $r=0$ to the system (\ref{log15}):
\begin{equation}
\begin{cases}
\displaystyle{\dot{\theta} = 2 (\sqrt{\mu} \cos{\psi} \sin{\theta} - \sin{\psi} \cos{\theta})}\\
\displaystyle{\dot{\psi} = 0.}
\end{cases}
\label{log16}
\end{equation}

We obtain a family of solutions $\psi = \psi_0 = const,$ whereas the vector field vanishes
along the deformed circles $\displaystyle{\{(\theta,
\psi)\hspace{0.1cm}| \hspace{0.1cm}\psi=\psi_0, \hspace{0.1cm}
\sqrt \mu \cos{\psi_0} \sin{\theta} - \sin{\psi_0} \cos{\theta} =
0\}}.$ Notice that for $\mu =1,$ the curves of equilibria transform into
circles given by $\displaystyle{\{(\theta,
\psi)\hspace{0.1cm}|\hspace{0.1cm} \psi=\psi_0,\hspace{0.1cm}
\theta=\psi_0, \hspace{0.2cm} \theta=\pi + \psi_0\}}.$ 
For the general case $\mu \neq 1$, let us denote by $C^+$ the equilibrium 
curve that passes  through ${\theta = \psi =0}$ and by $C^{-}$ the curve that 
passes through ${\theta = \psi =\pi}.$ It is immediate that $C^+$ is a repeller 
and $C^{-}$ is an attractor. The dynamical behavior on the collision manifold $\Lambda$
 is illustrated in Figure 1. Since $\dot\psi =0$, the flow follows the parallel 
lines $\psi=\psi_{0}= const.$

\vspace{1cm} Similarly, for $r=R_{max}$ we obtain another
invariant manifold $\Omega = \Omega_{h}$ for (\ref{log15}), namely the manifold
corresponding to the oval of zero velocity.
Recall that $R_{max}$ is the value which cancels $h r^2 +1 -f(r) =1 - f(r),$ ($h$ is set to be
zero). The flow on $\Omega$ is given by:

\begin{equation} \begin{cases}
\displaystyle{\dot{\theta} = 0}\\
\displaystyle{\dot{\psi} = R_{max}^{2}(\cos{\psi} \sin{\theta} -
\sqrt{\mu} \sin{\psi} \cos{\theta}).}
\end{cases}
\label{log17}
\end{equation}
The dynamical behavior on the zero-velocity manifold is similar to that on the collision manifold: 
the torus $(r,\theta, \psi) \in R_{max} \times \mathbb{S}^1 \times \mathbb{S}^1$
that represents $\Omega_{0}$ is covered by orbits parallel to
$\theta =\theta_{0} =const.$ There are again, two skewed circles of
equilibria, $ \{ (\theta, \psi) \hspace{0.1cm} | \hspace{0.1cm}
\theta = \theta_0, \hspace{0.1cm}  \cos{\psi} \sin{\theta} -
\sqrt{\mu} \sin{\psi} \cos{\theta}=0, \}.$ Denoting by $V^+$ the
curve that passes through $(0,0)$ and by $V^-$ the curve that passes through
$(0,\pi)$, it follows that $V^+$ is an attractor and $V^-$ is a
repeller (see Figure 2).

\section{Global Dynamics of the Axis-Symmetric System  ($\mu = 1$)}

We return now the full system (\ref{log15}) and, taking into
account that (at least in a prime analysis) the anisotropy is
given by values of $\mu$ close to 1 but greater than 1, we
define the parameter $\epsilon:=\sqrt \mu -1>0$, which we will later treat as a small perturbation to the isotropic system. We also choose to work with the relative angle $\phi:= \theta-\psi$, instead of the angle $\theta.$ Then, written in terms of $\epsilon$ and $(r, \phi,
\psi)$ and neglecting the terms of order $\epsilon^2$ and higher,
the system (\ref{log15}) becomes:

\begin{equation} \begin{cases}
\displaystyle{\dot{r} = - \frac{r^{3}}{r^2+2} 2(1 - f(r)) \big[ \cos \phi+ \epsilon \cos (\phi+\psi) \cos \psi\big] }\\
\displaystyle{\dot{\phi} = [2 (1- f(r))-r^2]\sin \phi + }\\
\displaystyle{ + \epsilon
[2 (1- f(r))\cos \psi \sin (\phi+\psi)+ r^2 \sin \psi \cos (\phi+\psi)] }\\
\displaystyle{\dot{\psi} = r^{2} [\sin \phi -\epsilon \sin \psi
\cos (\phi+\psi)]}
\end{cases}
\label{system}
\end{equation}

In the absence of the small perturbation $(\epsilon =0),$  that is in the axis-symmetric case $\mu=1,$ the above system reduces to:
\begin{equation} \begin{cases}
\displaystyle{\dot{r} = - \frac{r^{3}}{r^2+2} 2(1 - f(r))\cos \phi }\\
\displaystyle{\dot{\phi} = [2 (1- f(r))-r^2]\sin \phi }\\
\displaystyle{\dot{\psi} = r^{2} \sin \phi }
\end{cases}
\label{unp-system}
\end{equation}
with $(r, \phi, \psi)$ on the solid torus  $[0, R_{max}] \times
\mathbb{S}^1 \times \mathbb{S}^1.$  The equilibrium solutions form two
circles along $\phi=0$ and $\phi =\pi$ and are given by $(0, 0,
\psi_0)$ and $(0,\pi, \psi_0)$ where $\psi_0$ can be any value in
$[0, 2 \pi].$ 

\vspace{0.2cm}
There are four invariant manifolds:

\vspace{0.2cm} - the collision manifold (at $r=0$), on which the
dynamics is given by:
\begin{equation} \begin{cases}
\displaystyle{\dot{\phi} = 2 \sin \phi}\\
\displaystyle{\dot{\psi} = 0}
\end{cases}
\label{6.2}
\end{equation}

\vspace{0.2cm} - the zero velocity manifold (at $r=R_{max}$), on which
we have:
\begin{equation} \begin{cases}
\displaystyle{\dot{\phi} = -R_{max}^2\sin \phi}\\
\displaystyle{\dot{\psi} = R_{max}^{2} \sin \phi}
\end{cases}
\label{6.3}
\end{equation}
or just $\displaystyle{\frac{d \phi}{d \psi} = -1;}$

\vspace{0.2cm} - the "$\sin \phi =0$" manifolds (when $\phi =0$ or
$\phi=\pi$), with:
\begin{equation} \begin{cases}
\displaystyle{\dot{r} = - \frac{r^{3}}{r^2+2}
2(1 - f(r)), \hspace{0.1cm} \text{if}\hspace{0.1cm} \phi =0}\\
\displaystyle{\dot{r} = \frac{r^{3}}{r^2+2}
2(1 - f(r)), \hspace{0.1cm} \text{if}\hspace{0.1cm} \phi =\pi.}\\
\end{cases}
\label{6.4}
\end{equation}

Recall that in the unperturbed case the angular momentum is conserved. In this case (see Observation 1): 
\begin{equation}
\displaystyle{C(s)=g(r) 2(1-f(r))\sin \phi =C=const,}
\label{6.7}
\end{equation}
for all $r \in [0, R_{max}].$ It is easy to see that
\begin{equation}
\displaystyle{C=0 \Longleftrightarrow  \big[ r=0
\hspace{0.1cm}\text{or} \hspace{0.1cm} r =R_{max}
\hspace{0.1cm}\text{or} \hspace{0.1cm} \sin \phi =0 \big].}
\end{equation}
In other words, the angular momentum is null (i.e. the motion is
rectilinear) if and only if the orbits are either on the collision
manifold, or on the zero velocity manifold, or connecting the two.
From this perspective, we can denote the surfaces "$\sin \phi =0$" as manifolds
of zero angular momentum. Thus we have:

\vspace{0.3cm} \noindent PROPOSITION 6.1 In the axis-symmetric singular
logarithmic problem the only orbits reaching the collision are the rectilinear ones. 

\vspace{0.3cm} \noindent PROPOSITION 6.2  In the axis-symmetric singular
logarithmic problem all orbits with nonzero angular momentum are
bounded, they do not fall/eject into/from the source, and they do
not reach maximum distance with respect to the source.

\vspace{0.3cm} \noindent We return to the analysis of the
unperturbed system ({\ref{unp-system}}) and notice that the system
decouples in the sense that the first two equations are
independent of $\psi.$ Thus, if one solves the equations of $r$
and $\phi,$ then $\psi$ is obtained by replacing the expressions
for $r$ and $\phi$ into the third equation and then integrating. Therefore, 
a detailed qualitative analysis of the reduced system:

\begin{equation} \begin{cases}
\displaystyle{\dot{r} = - \frac{r^{3}}{r^2+2} 2(1 - f(r))\cos \phi, }\\
\displaystyle{\dot{\phi} = [2 (1- f(r))-r^2]\sin \phi }\\
\end{cases}
\label{red-system}
\end{equation}
is extremely useful, as it may be extended to the full $(r, \phi,
\psi)$ space by introducing the third coordinate $ \psi$ at the
end of our investigation.

The reduced system is relatively easy to describe. The motion
takes place on the cylinder $[0, R_{max}] \times \mathbb{S}^1.$
There are four degenerate saddle  equilibria located at 
$(0,0)$, $(0,\pi),$ $(R_{max},0),$ $(R_{max}, \pi),$  and two
centers at $(R_0, \pi/2)$ and $(R_0,3\pi/2),$ where $R_0$ is the solution 
of the equation $2(1-f(r)) -r^2=0.$ A direct computation shows that the
eigenvalues for the centers are given by
$\displaystyle{\lambda_{1,2}=\pm iR_0^2 \sqrt
{2(R_0^2+2)/(R_0+2)}}.$ Also, there are four invariant manifolds
$\{r=0\},$ $\{r=R_{max}\},$ $\{ \phi=0 \},$ and $\{\phi= \pi \},$
 forming two heteroclinic cycles connecting the
saddle equilibria (see Figure 3).

\vspace{0.5cm} We now lift the dynamics from the $(r, \phi)$ phase
space into the full $(r, \phi, \psi)$ solid torus  by taking into
consideration the third $\psi \in \mathbb{S}^1$ coordinate, as
well as the dynamics on the collision manifold, zero velocity
manifold and the $"\sin \phi =0"$ zero angular momentum manifolds.

The global flow, which takes place in  the solid torus $[0,
R_{max}] \times \mathbb{S}^1 \times \mathbb{S}^1$, can be seen in
Figure 4 and is represented as follows:

- the outside boundary corresponds to the collision manifold $\{r=0 \};$

- the interior boundary corresponds to $\{r=R_{max}\}.$ 
The centered circle of the torus was "blown up" artificially to
an inside torus such that the dynamics on the zero velocity
manifold can be seen (this is just a visual artifact and does not
modify the analysis);

- the motion takes place in between the exterior boundary $\{r=0
\}$ and the interior boundary $\{r=R_{max}\};$

- the surface of zero angular momentum $"\sin \phi =0"$ divides the
space into two distinct global invariant manifolds: one
corresponding to ${\phi \in (0, \pi)}$ or, equivalently, to motion
with positive angular momentum ($C>0$), and one corresponding to
${\phi \in (\pi, 2\pi)}$ or simply, to $C<0;$ the flow is symmetric
with respect to the horizontal plane $C=0.$ In this latter plane, the physical 
motion is rectilinear and represents orbits
that are ejecting from the collision manifold, reaching the
maximum distance at the zero velocity manifold and falling back on
the collision manifold.

- there are two periodic orbits situated symmetrically with
respect to the horizontal plane ($C=0$), namely: $\{(r, \phi, \psi)
\hspace{0.1cm}| \hspace{0.1cm} r= R_0, \hspace{0.1cm} \phi= \pi/2,
\hspace{0.1cm} \psi= \psi_o+R_o s; \hspace{0.1cm} s \geq 0,
\hspace{0.1cm} \psi_0 = \psi(0) \}$ and $\{(r, \phi, \psi)
\hspace{0.1cm}| \hspace{0.1cm} r= R_0, \hspace{0.1cm} \phi= 3
\pi/2, \hspace{0.1cm} \psi= \psi_o  - R_o s; \hspace{0.1cm} s \geq
0, \hspace{0.1cm} \psi_0 = \psi(0) \}.$ Around those two periodic
orbits, the phase-space is foliated by tori-like surfaces, parametrized by
the angular momentum integral (\ref{6.7});

- the equilibria on the collision manifold are connected with the equilibria on
the zero velocity manifold, through heteroclinic
cycles of the form:

$$ a \longrightarrow b \longrightarrow c \longrightarrow d \longrightarrow e \longrightarrow f \longrightarrow g \longrightarrow h \longrightarrow a,$$

\noindent where $a = (0, 0, \psi_0),$ $b = (0, \pi, \psi_0),$ $c = (R_{max}, \pi, \psi_0),$ $d = (R_{max}, 0, \psi_0+\pi),$ $e = (0, 0, \psi_0 +\pi),$ $f = (0, \pi, \psi_0 +\pi),$ $g= (R_{max},\pi, \psi_0+\pi)$ and $h = (R_{max}, 0,
\psi_0+2\pi) \equiv (R_{max}, 0, \psi_0),$  for each fixed $\psi_0$ fixed in $[0, 2\pi]$ corresponding one cycle.

\vspace{1cm} 
Let us look at the surfaces with positive values of 
the angular momentum, $C > 0$ (for negative $C,$ by symmetry, the flow
is identical but of opposite sense). Since each value $C=const$ represents a tori-like surface, the space 
contains a series of invariant manifolds nested one into the other. $C$ varies from $0$ to a
maximal value corresponding to the degenerate torus $r=R_0,$ {\it i.e.} a
circle (see Figure 5). 

The dynamics will change as one shifts from the tori close to $C=0$ towards the tori near 
the periodic orbit $r=R_0.$ In the $(\phi, \psi)$ plane, the orbits close to the collision manifold 
$\{ r = 0 \}$ have very large slopes $d\phi / d\psi$ (for example, the orbits on $c_1$ in Figure 5). 
The orbits close to the periodic orbit $r=R_0$ have slopes close to $-1$ 
(e.g. the orbits on $c_3$ in Figure 5). In between, there is a smooth transition in the 
values of the slopes (one has to keep in mind that the flow is continuous and therefore orbits which are initially close have to stay close at all times).

For a fixed value of $C$ or, in other words, for a fixed torus, the
motion is bounded between a minimum and a maximum value, which
depend directly on $C.$ We note that in the real physical space $(x_1, x_2),$
this translates into a bounded motion between a minimum and a maximum value, representing a
loop orbit. The periodic solutions, $(R_{0},\pi/2,\psi)$ and $(R_{0},3\pi/2,\psi),$ 
correspond to the two parent families of the loop orbits, one counter-clockwise $(C>0)$ and 
the other clockwise $(C<0)$ \cite{BT}. We shall discuss the family of loop 
orbits in more detail in Sections 7 and 8.

\section{Dynamics of the Non Axis-Symmetric System ($\mu \neq 1$)}

In this section, we proceed to investigate the general form of the non axis-symmetric system ({\ref{system}}). We will limit our analysis to the case in which the anisotropy is small ($\mu \simeq 1$) and therefore we can treat it as a small perturbation $(\epsilon=\sqrt{\mu}-1>0)$ to the isotropic system.

In the following, we derive the curves of equilibria on the collision manifold and on the 
zero velocity manifold. We observe that these curves, located in the horizontal plane $(C=0),$ remain upon the perturbation within $O(\epsilon)$ distance of this plane. We then prove that most
orbits in the perturbed case avoid the origin, a result initially discovered in the numerical study of Miralda-Escud\'e $\&$ Schwarzschild \cite{MS}. For this, we show that the curves of equilibria on the collision manifold are degenerate saddles, {\it i.e.} the equilibrium points on these curves admit a 2D 
stable and a 2D unstable manifold. The implication is that, in the three-dimensional space, the dimension of the set of initial conditions leading to collision is two ({\it i.e.} the Lebesgue measure of the set is zero). In physical space this translates into a zero probability of finding orbits falling into the source.

\vspace{0.5cm} \noindent LEMMA 7.1 The equilibria of the vector
field (\ref{system}) consists of four closed curves, two belonging
to the collision manifold and the other two to the zero velocity
manifold. The collision manifolds curves of equilibria are given
by:
 $$ C_1:=\{(r, \phi, \psi) \hspace{0.1cm} | \hspace{0.1cm} r=0,
\hspace{0.1cm} \phi = \arctan[(1- \epsilon) \tan \psi] - \psi,
\hspace{0.1cm}
 \psi=\psi_0 \in [0, 2\pi)\hspace{0.1cm} \} \cup $$
 \begin{equation} \cup (0, 0, \pi/2) \cup (0, 0, 3\pi/2)\label{r=0equil1}\end{equation}
 and
 $$ C_2:=\{(r, \phi, \psi) \hspace{0.1cm} | \hspace{0.1cm} r=0,
\hspace{0.1cm} \phi = \pi + \arctan[(1- \epsilon) \tan \psi] -
\psi, \hspace{0.1cm}
 \psi=\psi_0 \in [0, 2\pi)\hspace{0.1cm} \} \cup $$
\begin{equation} \cup (0, \pi, \pi/2) \cup (0, \pi, 3\pi/2).\label{r=0equil2}\end{equation}
The curves of equilibria  on the zero velocity manifold are given
by: $$Z_1:=\{(r, \phi, \psi) \hspace{0.1cm} | \hspace{0.1cm}
r=R_{max}, \hspace{0.1cm} \phi = \arctan[(1+ \epsilon) \tan \psi]
- \psi ,\hspace{0.1cm}
 \psi=\psi_0 \in [0, 2\pi) \hspace{0.1cm} \} \cup $$
\begin{equation}(R_{max}, 0, \pi/2) \cup (R_{max}, 0, 3\pi/2), \label{r=Requil1}\end{equation}
and $$Z_2:=\{(r, \phi, \psi) \hspace{0.1cm} | \hspace{0.1cm}
r=R_{max}, \hspace{0.1cm} \phi = \pi +\arctan[(1+ \epsilon) \tan
\psi] - \psi ,\hspace{0.1cm}
 \psi=\psi_0 \in [0, 2\pi) \hspace{0.1cm} \} \cup $$
 \begin{equation} (R_{max}, \pi, \pi/2) \cup (R_{max}, \pi, 3\pi/2). \label{r=Requil2}\end{equation}

\vspace{0.5cm}\noindent PROOF The proof follows by inspection. One
can verify directly that the two families of curves cancel the
vector field. To see that there are no other equilibria, suppose
$r \neq 0$ and $r \neq R_{max}.$ Then $\dot{r} =0 $ and $\dot
{\psi}=0$ lead to
$$
\begin{cases}
(1+\epsilon)\cos (\phi+\psi)\cos\psi = -  \sin
(\phi+\psi)\sin\psi\\
(1+\epsilon)\cos (\phi+\psi)\sin\psi = \sin
(\phi+\psi)\cos\psi.
\end{cases}
$$ It is easy to check that the above relations cannot coexist.

\vspace{0.5cm} \noindent OBSERVATION 7.2 $ C_1$ and $C_2$ are
within $O(\epsilon)$ distance of the circles $$\displaystyle{\{(r,
\phi, \psi) \hspace{0.1cm} | \hspace{0.1cm} r=0, \hspace{0.1cm}
\phi = 0, \hspace{0.1cm} \psi=\psi_0 \in [0, 2\pi)\hspace{0.1cm}
\}}$$ and respectively, $$\displaystyle{\{(r, \phi, \psi) \hspace{0.1cm} |
\hspace{0.1cm} r=0, \hspace{0.1cm} \phi = \pi, \hspace{0.1cm}
\psi=\psi_0 \in [0, 2\pi)\hspace{0.1cm} \}.}$$ Also,
on the zero velocity manifold, $ Z_1$ and $Z_2$ are within
$O(\epsilon)$ distance of the circles $$\displaystyle{\{(r, \phi,
\psi) \hspace{0.1cm} | \hspace{0.1cm} r=R_{max}, \hspace{0.1cm}
\phi = 0, \hspace{0.1cm} \psi=\psi_0 \in [0, 2\pi)\hspace{0.1cm}
\}}$$ and $$\displaystyle{\{(r, \phi, \psi) \hspace{0.1cm} |
\hspace{0.1cm} r=R_{max}, \hspace{0.1cm} \phi = \pi,
\hspace{0.1cm} \psi=\psi_0 \in [0, 2\pi)\hspace{0.1cm} \},}$$
respectively.

\vspace{0.5cm} \noindent LEMMA 7.3 The curves of equilibria $Z_1$
and $Z_2$ on the zero velocity manifold are degenerate saddles
(see Figure 6). More precisely, for a fixed $\psi_0,$ the two
corresponding equilibria
$$( R_{max},\arctan[(1+ \epsilon) \tan \psi_0] -
\psi_0, \psi_0) \in Z_1$$ and
$$( R_{max},\pi+ \arctan[(1+ \epsilon) \tan \psi_0] - \psi_0,
\psi_0) \in Z_2$$ behave like saddles in the $(r, \phi)$ plane.

\vspace{0.1cm}\noindent PROOF Recall from Section 6 that on the
zero velocity manifold the flow is degenerate  and in the $(\phi,
\psi)$ coordinates it reads:
$$\begin{cases} \dot{\phi} = - R_{max}^2 [\sin (\phi+\psi)\cos \psi -
(1+\epsilon) \cos(\phi +\psi)\sin \psi],\\
\dot{\psi }= R_{max}^2 [\sin (\phi +\psi) \cos \psi -(1+ \epsilon
)\cos(\phi +\psi)\sin \psi].\end{cases}
$$

\vspace{0.3cm} By symmetry, it is sufficient to study the flow around
one of the curves of equilibria, let's say $Z_1$ (around $Z_2$ we
just have to reverse the arrows). For simplicity, for a fixed
$\psi_0,$ we denote the $\phi$ component of an equilibria by $\phi
_Z(\psi_0),$ {\it i.e.} $\phi _Z(\psi_0): = \arctan[(1+ \epsilon) \tan
\psi_0] - \psi_0.$ Observe that $(R_{max}, \phi_Z(\psi_0),
\psi_0)$ are simple roots for $\dot{r}$ and $\dot{\phi}.$
Computing the eigenvalues at $(R_{max}, \phi_Z(\psi_0), \psi_0),$
we obtain:
$$\begin{cases}\lambda_r= -\frac{R_{max}^{3}}{R_{max}^2+2}\Big(-
2f'(R_{max})\Big)
 \big[ \cos \phi_Z(\psi_0)+ \epsilon \cos (\phi_Z (\psi_0)+\psi_0) \cos \psi_0\big]\\
\lambda_{\phi} =- 2 \cos \big(\phi_Z (\psi_0)+\psi_0\big) \cos
\psi_0 -
\epsilon \sin \big(\phi_Z( \psi_0)+\psi_0\big) \sin \psi_0\\
\lambda_{\psi} =0.\end{cases}$$ Recall that $\psi_0$ is fixed in $[0,
\pi/2].$ If $\psi_0 = \pi/2,$ we obtain $\phi_Z (\psi_0) = \phi
(\pi/2) = 0$ and

$$\begin{cases}\displaystyle{\lambda_r= -\frac{R_{max}^{3}}{R_{max}^2+2}(-
2f'(R_{max}) >0,}\\
\lambda_{\phi} =
- \epsilon <0,\\
\lambda_{\psi} =0.\end{cases}$$ Therefore $(R_{max}, 0, \pi/2)$ is
a degenerate saddle. If $\psi_0  \in [0, \pi/2),$ we have:

$$\begin{cases}\displaystyle{\lambda_r= -(1+\epsilon)\frac{R_{max}^{3}}{R_{max}^2+2}(-
2f'(R_{max})) \frac{\cos \big(\phi_Z (\psi_0)+\psi_0\big)}{\cos \psi_0},}\\
\displaystyle{\lambda_{\phi} = - (2\cos^2 \psi_0 + \epsilon \sin^2
\psi_0) \frac{\cos \big(\phi_Z (\psi_0)+\psi_0\big)
}{\cos \psi_0},}\\
\lambda_{\psi} =0.\end{cases}$$ Since $\psi_0 \in [0, \pi/2)$ and
$\phi _Z(\psi_0) = \arctan[(1+ \epsilon) \tan \psi_0] - \psi_0,$
it results that  $\phi _Z(\psi_0 ) + \psi_0  \in [0, \pi/2)$ and
furthermore, $\cos \big(\phi _Z(\psi_0)+\psi_0\big)>0.$ Therefore
$\lambda_r>0$ and $\lambda_{\phi}<0.$

 The other cases where $\psi_0 \in (\pi/2, \pi) \cup [\pi, 3\pi/2] \cup (3\pi/2, 2\pi)$
 can be treated similarly, reaching the same conclusion, {\it i.e.} the
 equilibria points $Z_1$ are all
 degenerate saddles, with:
 $$\begin{cases}\displaystyle{\lambda_r >0,}\\
\lambda_{\phi} <0,\\
\lambda_{\psi} =0.\end{cases}$$

\vspace{0.5cm} \noindent LEMMA 7.4 The curves of equilibria $C_1$ and $C_2$ on the
collision manifold are degenerate saddles (see Figure 6). More
precisely, for a fixed $\psi_0,$ the two corresponding equilibria
$$(0, \arctan[(1- \epsilon) \tan \psi_0] - \psi_0, \psi_0),$$
$$(0, \pi + \arctan[(1- \epsilon) \tan \psi_0] - \psi_0,
\psi_0)$$  behave like degenerate saddles.

\vspace{0.5cm}\noindent PROOF The equations of the flow on the
collision manifold are
$$\begin{cases}
\dot{\phi} = 2[(1+\epsilon)\sin (\phi +\psi)
\cos \psi - \cos(\phi +\psi)\sin \psi] \\
\dot{\psi} =0\end{cases}$$

As in the previous proof, it is sufficient to investigate the flow
around the equilibria curve $C_1$ (around $C_2,$ by symmetry, we
just have to reverse the arrows). Fixing $\psi_0 \in [0, 2\pi),$ we
denote the corresponding fixed point on $C_1$ by $(0,
\phi_C(\psi_0), \psi_0 )$ and proceed to calculate its
eigenvalues. We obtain:
$$\begin{cases}\displaystyle{\lambda_r =0,}\\
\lambda_{\phi} = 2\Big[(1+\epsilon) \cos \big(\phi_C
(\psi_0)+\psi_0\big) \cos \psi_0 + \sin \big(\phi_C
(\psi_0)+\psi_0\big) \sin \psi_0\Big],\\
\lambda_{\psi} =0.\end{cases}$$ We observe that the  vector field
({\ref{system}}) manifests a degeneracy around $r=0.$ This means that any fixed
point on the curve $C_1$ might have a non-hyperbolic character in the $(r, \phi)$
plane. For the moment, let us discuss the sign of
$\lambda_{\phi}.$ Similar to the analysis in the case of the curve $Z_1$, we fix $\psi_0 \in [0,
\pi/2].$ If $\psi_0 =\pi/2,$ we have $\phi_C (\pi/2) = 0$ and
therefore $\lambda_{\phi} =2>0.$ If $\psi_0 \in [0, \pi/2),$ we
obtain:
$$\lambda_{\phi} = \frac{\sin(\psi_0)}{(\phi_C
(\psi_0)+\psi_0\big)}.
$$ Since $(\phi_C
(\psi_0)+\psi_0\big) = \arctan \big[(1-\epsilon)\tan \psi_0 \big]
\in (0, \pi/2),$ it follows that $\sin(\phi_C (\psi_0)+\psi_0\big)
>0 $ and therefore $\lambda_{\phi}>0.$ The same type of reasoning
applies for all the other cases, {\it i.e.} for $\psi_0\in (\pi/2, \pi)
\cup [\pi, 3\pi/2] \cup (3\pi/2, 2\pi).$ Therefore we have proved that
$\lambda_\phi >0 \hspace{0.3cm} \hbox{for any fixed} \hspace{0.2cm}
\psi_0 \in [0, 2\pi).$

In conclusion, the flow around the equilibria $(0,\phi_C (\psi_0),
\psi_0 ) \in C_1$ is degenerate, with eigenvalues:
$$\begin{cases}\displaystyle{\lambda_r =0}\\
\lambda_{\phi}  >0\\
\lambda_{\psi} =0.\end{cases}$$

Obviously, the linear approximation of the flow does not provide
enough information about the behavior around the fixed points. Let
us take a closer look at the vector field around $(0,\phi_C
(\psi_0), \psi_0 ) \in C_1.$

Notice that on the $\psi$ direction the flow is null,
as every point on the circle $\psi\in [0, 2\pi)$ is a parameter
for a fixed point. It remains that in order to describe the
asymptotical behavior around $(0,\phi_C (\psi_0), \psi_0 ),$ one
has to investigate the flow in the $(r, \phi)$ coordinates.
$\lambda_r =0$ generates the center manifold $E^c$ --the span of
the zero eigenvector (\cite{GH}, \cite{W}).
The general theory ensures the existence of an invariant manifold
$W^c$ tangent to $E^c$ at $(r, \phi_C(\psi_0)).$ $W^c$ may not be
unique and, usually, it involves a loss of smoothness. Also, around
the fixed point, $W^c$ is described by a one-parameter family of
curves, {\it i.e.} $W_c = \{(r, \phi) \hspace{0.2cm} |\hspace{0.2cm}
\phi = \phi_{\beta}(r), \beta \in \mathbb{R} \}.$

In our case, we will compute $W^c$ directly near $(r,
\phi_C(\psi_0)),$ as follows (see for example, \cite{K}): We know that by a proper
transformation of coordinates the equations for $(r, \phi)$ have
the structure
\begin{equation}
\begin{cases}
  \dot{r} = - \displaystyle{\frac{r^{3}}{r^2+2} 2(1 - f(r)) F_1(\phi, \psi; \epsilon) }\\
  \dot{\phi} = \lambda_{\phi}(\phi - \phi_C(\psi_0)) + \ldots .\\
\end{cases}\label{precenter}
\end{equation}
Near $r=0,$ we can make the approximation:
\begin{equation}\displaystyle{\frac{r^{3}}{r^2+2} 2(1 - f(r))=
r^{3} \frac{1}{2}(1+r^2/2+ \ldots) 2(1 - f(r)) =
}\label{expansion}
\end{equation}
$$=r^3 + higher \hspace{0.2cm} order \hspace{0.2cm} terms$$ (note that we are not expanding $f(r),$
a function that is only differentiable, but we are merely looking for
the dominant term as $r$ goes to zero). Therefore, around $(r,
\phi_C(\psi_0))$ we have:
\begin{equation}
\begin{cases}
  \dot{r} = - ar^3  + \ldots ,\\
  \dot{\phi} = \lambda_{\phi}(\phi - \phi_C(\psi_0)) + \ldots ,\\
\end{cases}
\label{center}
\end{equation}
where $a: = F_1(\phi_C(\psi_0), \psi_0; \epsilon)$ is a positive
number. We obtain immediately that:
\begin{equation}
\displaystyle{\frac{dr}{d\phi}  =-\frac{ar^3}{\lambda_{\phi}(\phi
- \phi_C(\psi_0))}}, \label{postcenter}
\end{equation}
with solutions:
\begin{equation}
\displaystyle{\phi_{\beta} (r)= \left\{\begin{array}{c}
 \displaystyle{\beta e^{{\lambda_{\phi}}/(2r^2)} \hspace{0.4cm}\hbox{if} \hspace{0.2cm}r>0 }\\
  0 \hspace{1.7cm} \hbox{if}\hspace{0.2cm} r=0,
\end{array}
\right.}
 \label{family}
\end{equation} where $\beta \in \mathbb{R}$ is a parameter.

We sketch the family of solutions $\phi_{\beta}(r)$ in Figure 7. As it can be
easily seen, the fixed point $(0, \phi_C(\psi_0))$ is indeed a
saddle.

\vspace{1cm} Since the above reasoning applies for any of the
points of equilibria, and using the Lemmas 7.1 and 7.3, we can state the following:

 \vspace{0.5cm} \noindent THEOREM 7.5 Let be the system
(\ref{system}). Then the equilibria are given by the curves $C_1,$
$C_2$ $Z_1$ and $Z_2,$ as defined in Lemma 7.1, and each of this
curves admits a two-dimensional stable manifold and a two-dimensional 
unstable manifold.

 \vspace{0.5cm} \noindent COROLLARY 7.6 In the anisotropic logarithmic problem
 the Lebesgue measure of the set of initial conditions that lead to collision is zero.

\vspace{0.5cm} The last result states that in the phase space the
probability of choosing any initial conditions that lead to collision
is zero. The set of such initial conditions is formed by a two-dimensional
 manifold embedded in a three-dimensional phase space.
Since we reach a similar conclusion in the isotropic case, we can
conclude that for the singular logarithmic potential the anisotropy
does not increase the probability of finding orbits falling
into the source. In other words, we have proved analytically that
in the non axis-symmetric case all orbits (with the exception of a set with measure zero)
are centrophobic, a result that was also noticed in previous numerical
studies \cite{MS}.

\clearpage We turn our attention now to the behavior of the
angular momentum $C(s).$ Recall that $C(s)$ was given in
({\ref{moment}}). For $h=0$ and in terms of $\epsilon$ and $(r,
\phi, \psi),$ the relation ({\ref{moment}}) becomes:
\begin{equation}
\displaystyle{ C(s) =  g(r)2[1- f(r) ] \Big[(1+\epsilon)\sin
\big(\phi+\psi\big) \cos \psi - \cos(\phi+\psi\big) \sin \psi
\Big].} \label{momenteps}
\end{equation}

We note that under the perturbation, the horizontal surface $C=0$ (see Figure 4) splits up,
 and some of its vestiges are to be found along the two-dimensional stable manifold of $C_1$ and 
along the unstable manifold of $C_2.$ In physical space, this corresponds to the case of
rectilinear orbits (C=0), which are falling into or ejecting from the source.
 
The variation of angular momentum (\ref{angular}) reads in our
coordinates:
\begin{equation}
\displaystyle{ \frac{dC}{ds} = - \epsilon  g(r)2[1- f(r)
]^{3/2}\sin \psi \cos \psi.} \label{momenteps}
\end{equation}

Since the derivative of $C$ is bounded, there are no "blow-up" type effects in the evolution of
$C.$ Also, since the product $g(r)\cdot 2[1- f(r)]$ is always positive, it follows
that the critical points of $C(s)$ correspond to $\psi \in \{ 0,
\pi/2, \pi, 3\pi/2 \}$ (see Figure 8). These orbits, for which $C(s)$ displays a sinusoidal-type behavior and 
admits four critical points, represent the family of loop orbits (see also a similar result in the study of Touma $\&$ Tremaine \cite{TT}).
 
On the other hand, there are orbits which, under perturbation, will slip in between the stable
manifold of $C_1$ and the unstable manifold of $C_2,$ switching the sign of the angular momentum. By the nature of the phase-space, which is a solid torus, these orbits must wind around 
indefinitely. These orbits pertain to the family of "boxlet" orbits discovered by Miralda-Escud\'e $\&$ Schwarzschild \cite{MS}. For large perturbations, more and more orbits will break away from the curves of equilibria $C_1$ and $C_2,$ and become boxlet, a result which was also pointed out in \cite{MS}. The separatrix that divides
the phase-space between loop and boxlet orbits intersects the $(C=0)$ plane at $\psi = \pi/2$ and 
$\psi =3\pi/2$ (see \cite{TT}). Some of the orbits that wind up inside the torus will eventually close, becoming the parent orbits of the resonant families ({\it i.e.} banana, fish, pretzel, etc) \cite{MS}.

\vspace{0.3cm}
We do not present here a rigorous proof for the existence of the boxlet obits or the resonances, 
leaving it for a future study. However, we present below a partial result concerning the orbits which preserve the sign of the angular momentum.

 Let us notice that the angular momentum can be regarded
as a function of two arguments, namely as $C=C(s,\epsilon).$ Then, 
around the equilibrium point $(s,0),$ we have the following approximation:

\begin{equation}
C(s, \epsilon) = C(s, 0)+ \epsilon \frac{\partial C(s,
\epsilon)}{\partial \epsilon}\Bigg |_{\epsilon=0}+ O(\epsilon^2)
\label{cepsilon}
\end{equation}
or, in our coordinates,
\begin{equation}
C(s, \epsilon) = C(s, 0)+ \epsilon g(r)2[1- f(r) ]
\sin\big(\phi+\psi\big) \cos \psi. \label{cexp}
\end{equation}
But $C(s, 0)$ is a constant, since this is the case of the unperturbed
motion, where the angular momentum is an integral of motion. Denoting $C_0:=C(s,
0),$ it follows that:
\begin{equation}
C(s, \epsilon) = C_0+ \epsilon g(r)2[1- f(r) ] \sin(\phi+\psi\big)
\cos \psi. \label{cexpo}
\end{equation}
Since $\displaystyle{\epsilon g(r)2[1- f(r) ]
\sin\big(\phi+\psi\big) \cos \psi}$ is bounded, we conclude that
$C(s)$ preserves its sign for large values of $C_0$ and
small $\epsilon.$ This is best noticed near the unperturbed
periodic orbits, for example, $(r, \phi, \psi )=(R_0, \pi/2, \psi ),$ $\psi \in
[0, 2\pi),$ where $C_0 =$
$\displaystyle{e^{-1/R_0^2}2(1-R_0^2 \ln
R_0)}$ and where the perturbation induces loop orbits (see
Figure 9).

\section{The Orbital Structure}

The orbital structure can also be retrieved numerically, by integrating the general form of the system (\ref{system}). The return to the initial coordinates in the physical space, $(x_1,x_2),$ can then be made through the relations:

\begin {equation}
 \begin{cases}
x_1= q_1 / \mu =(1/\mu)\cdot r e^{(-1/r^2)}\cos (\phi + \psi)\\
x_2= q_2 =r e^{(-1/r^2)}\sin (\phi + \psi),
\end{cases}
\label{x si y}
\end{equation}

\noindent and recalling that the time scale was modified such that both the
singularity and the zero velocity manifold are now reached in an infinite
time.

By appropriately choosing the initial conditions and recalling that $\epsilon = \sqrt{\mu} -1 = 1/b - 1,$
one can retrieve the resonance families. For example, we know that the family of loop orbits (1:1) develops 
around the two periodic orbits $r=R_0$ (see Figure 4), {\it i.e.} near:
\begin {equation}
 \begin{cases}
x_1= (1/\mu)\cdot r e^{(-1/r^2)} \cos (\pi/2 + \psi),\\
x_2= r e^{(-1/r^2)} \sin (\pi/2 + \psi),\\
\psi = R_0^2 s + \psi_0.
\end{cases}
\end{equation}

and 
\begin {equation}
 \begin{cases}
x_1= (1/\mu)\cdot r e^{(-1/r^2)} \cos (3\pi/2 + \psi),\\
x_2= r e^{(-1/r^2)} \sin (3\pi/2 + \psi),\\
\psi = - R_0^2 s + \psi_0.
\end{cases}
\end{equation}

\noindent The two periodic orbits correspond to the two parent orbits of the loop family, one evolving clockwise ($C<0$), the other counter-clockwise ($C>0$). Integrating the system (\ref{system}) for the case $\epsilon = 0$, around either $(R_{0}, \pi/2, \psi)$ or $(R_{0}, 3\pi/2, \psi),$ one retrieves a circle of radius $R_{0}$, as indicated with thick line in Figure 9. The same figure shows, with dotted line, the orbital structure in the vicinity of the parent orbit for the case when $\epsilon =0.1$ (corresponding to $b=0.9$).

\vspace{0.3cm}
Besides the loop resonance, several other resonances can be recovered from the new system of differential equations. We remark that under large perturbations ({\it i.e.} large deviations from axis-symmetry), the families of minor orbits will occupy more and more of the phase-space. We do not attempt here to cover the entire phase-space, but rather give a few examples of families of minor orbits found with our new system of equations. For example, Figure 10 shows an example of an orbit around the fish (3:2) resonance, in the case $\epsilon=0.3$ (corresponding to $b=0.7$) and the initial conditions $r=0.9,$ $\phi = \pi/12$ and $\psi=\pi/8.$ Figure 11 shows an orbit around the pretzel (4:3) resonance, when $\epsilon =0.1$ and $r=0.45,$ $\phi = 0,$ $\psi=\pi/14.$ Figure 12 shows an orbit near the (5:3) resonance, 
obtained for $\epsilon=0.3,$ $r=0.4,$ $\phi = \pi/12,$ $\psi=\pi/6.$ We note however, that these parameters should be taken only as a guidance for the location of the resonances.

\section{Conclusions}

Previous numerical studies have revealed several important aspects of the orbital structure of the logarithmic potential: the division between loop and box orbits, the presence of resonances, the scattering effect of the singularity (which renders the box orbits unstable), and the transition to chaos. 
We have performed an analytical study of the singular logarithmic potential and proved several of these results. 

We summarize our results as follows:

- we provide a description of the dynamics near the singularity and at the maximum distance from the source permitted for a given level of energy.

- in the axis-symmetric case, we retrieve the complete global dynamics of the orbits and describe it on a solid torus bounded by the two surfaces $ \{ r=0 \}$ and $ \{ r=R_{max} \}$. We find analytically the two periodic orbits $r=R_0$, which correspond in physical space to the two parent families of the loop orbits (in clockwise and respectively, in counter-clockwise motion).

- in the non axis-symmetric case, we prove that all orbits, except a negligible set, are centrophobic --a
result that has been originally discovered in the numerical study of Miralda-Escud\'e $\&$ Schwarzschild \cite{MS}.

- in the same non axis-symmetric case, we show that there exist orbits which preserve the sign of the angular momentum and retrieve the loop resonance. Finally, we also show how several other minor family orbits can be obtained from the our new system of equations.

The analytical description in general non axis-symmetric case remains still open, several problems requiring further investigation. One of them is the retrieval of the family of boxlet orbits, which are known to dominate the dynamics near the singularity \cite{MS}.
We conclude, however, that under large perturbations (large deviations from axis-symmetry), most of the phase space is occupied by orbits that slip in between the stable manifold of $C_1$ and the unstable manifold of $C_2$ and wind around indefinitely. Most of these orbits pertain to the family of the boxlet orbits. 
Some of these winding orbits may be closed and would lie at the origin of the resonant families observed experimentally. Further work in this direction will have to include an in-depth analysis of resonances in the $(x_1, x_2)$ variables and a Fourier expansion of the periodic solutions for $r(s).$

Another aspect that remains to be clarified is the existence of the stochastic orbits near the singularity and the transition to chaos. The analytical study in this case is hindered by the fact that the system given by the Hamiltonian (\ref{log2}) does not admit any hyperbolic equilibrium points, and therefore, the perturbative methods which are usually employed in proving the existence of chaos (including the Melnikov method \cite{M}) become unapplicable. We note that the Melnikov method has been used in the past in the case of the logarithmic potential \cite{G}, however not on the exact Hamiltonian, but on the integrable St\"ackel Hamiltonian \cite{S}. Understanding the onset of chaos in the singular logarithmic potential is a very important problem for the construction of galaxy models based on libraries of orbits (see for example, \cite{Sc}) - in which generally, it is a priori assumed that the stochastic orbits play a negligible role.

\newpage
{\small Figure 1: On the collision manifold, the flow follows the parallel 
lines $\psi=\psi_{0}= const.$ The flow vanishes on the equilibrium curves $C^{+}$ and $C^{-}.$}

{\small Figure 2: On the zero velocity manifold, the flow follows the parallel  lines $\theta=\theta_{0}= const.$ The flow vanishes on the equilibrium curves $V^{+}$ and $V^{-}.$}

{\small Figure 3: The reduced phase-space ($r,\phi$) in the case $\mu=1.$
There are six equilibrium points: two centers, $(R_0, \pi/2)$ and $(R_0,3\pi/2),$ and four saddle equilibria, $(0,0)$, $(0,\pi),$ $(R_{max},0),$ $(R_{max}, \pi).$}

{\small Figure 4: The global flow in the axis-symmetric case ($\mu=1$),
described in a solid torus delineated by the two surfaces $\{r=0 \}$ and $\{ r=R_{max} \}$. 
The full lines denote the heteroclinic cycle for the case $\psi_0 = 0$.
The horizontal plane ($C=0$) divides the phase-space in two symmetric invariant subspaces (see text for details).}

{\small Figure 5: Section on the Solid Torus at $\psi=\psi_0.$ Around the two periodic orbits $r=R_0$, the phase-space is foliated by tori-like surfaces of constant angular momentum, $C$.}

{\small Figure 6: The curves of equilibria $Z_1$ and $C_1$, located on the collision manifold and on the zero-velocity manifold, respectively. Each equilibrium point on these two curves behaves like a degenerate saddle.}

{\small Figure 7: Curves in the $\phi_{\beta} (r)$ family around the saddle point $(0, \phi_C(\psi_0)).$}

{\small Figure 8: The curves of equilibria $C_1$, $C_2$, $Z_1$, $Z_2$ under the perturbation $\epsilon.$ The arrows show the behavior of the flow in their neighborhood. The manifolds $C_1$ and $C_2$ intersect transversely along the lines $\sin \phi =0$ and $\cos\phi=0.$}

{\small Figure 9: The circle $r=R_0$ (thick line) represents one of the parent families of loop orbits. The dotted line shows the orbital structure that is triggered by applying a perturbation ($\epsilon =0.1$), to the periodic orbit (shown here is the integration only over a finite time).}

{\small Figure 10: Orbit around the fish (3:2) resonance.} 

{\small Figure 11: Orbit around the pretzel (4:3) resonance.} 

{\small Figure 12: Orbit around the (5:3) resonance.} 

\end{document}